\documentclass[conference]{IEEEtran}
\usepackage{epsfig}
\usepackage{times}
\usepackage{float}
\usepackage{afterpage}
\usepackage{amsmath}
\usepackage{amstext,cite}
\usepackage{amssymb,bm}
\usepackage{latexsym}
\usepackage{color}
\usepackage{graphicx}
\usepackage{amsmath}
\usepackage{amsthm}
\usepackage{graphicx}
\usepackage[center]{caption}
\usepackage{pstricks}
\usepackage{caption}
\usepackage{subcaption}
\usepackage{booktabs}
\usepackage{multicol}
\usepackage{lipsum}
 \usepackage[T1]{fontenc}
\usepackage{mathrsfs}
\usepackage[normalem]{ulem}

\newcommand{\Ac}{\mathcal{A}}
\newcommand{\Bc}{\mathcal{B}}
\newcommand{\Dc}{\mathcal{D}}
\newcommand{\Jc}{\mathcal{J}}
\newcommand{\Kc}{\mathcal{K}}
\newcommand{\Wc}{\mathcal{W}}
\newcommand{\Sc}{\mathcal{S}}
\newcommand{\Pc}{\mathcal{P}}

\allowdisplaybreaks

\newtheorem{thm}{Theorem}
\newtheorem{cor}{Corollary}

\newtheorem{example}{Example}

\providecommand{\definitionname}{Definition}

\begin{document}

\title{A Novel Index Coding Scheme and 
\\
its Application to Coded Caching}

\author{%
\IEEEauthorblockN{Kai~Wan}%
\IEEEauthorblockA{Laboratoire des Signaux et Systèmes (L2S)\\
CentraleSup\`elec-CNRS-Universit\`e Paris-Sud\\
91192 Gif-sur-Yvette, France\\
Email: kai.wan@u-psud.fr}%
\and
\IEEEauthorblockN{Daniela~Tuninetti}%
\IEEEauthorblockA{NICEST Laboratory\\
University of Illinois at Chicago\\
Chicago, IL 60607, USA\\
Email: danielat@uic.edu}%
\and
\IEEEauthorblockN{Pablo~Piantanida}%
\IEEEauthorblockA{Laboratoire des Signaux et Systèmes (L2S) \\
CentraleSup\`elec-CNRS-Universit\`e Paris-Sud\\
91192 Gif-sur-Yvette, France\\
Email: pablo.piantanida@centralesupelec.fr}%
}

\maketitle

\begin{abstract}
This paper 
proposes a novel achievable scheme for the index problem and 
applies it to the caching problem. 
Index coding and caching are noiseless broadcast channel problems where 
receivers have message side information.
In the index coding problem the side information sets are fixed, while 
in the caching problem the side information sets correspond the cache contents, which are under the control of the system designer. 
The proposed 
index coding scheme, based on distributed source coding and non-unique decoding,
is shown to strictly 
enlarge the rate region achievable by composite coding.
%
The novel index coding scheme applied to the caching problem 
is then shown to match an outer bound (previously proposed by the authors and also based on known results for the index coding problem) under the assumption of uncoded cache placement/prefetching.
\end{abstract}

\IEEEpeerreviewmaketitle{}

\section{Introduction}
\label{sec:intro}

The index coding problem, originally proposed by Birk and Kol in~\cite{birk1998informedsource}, is a distributed source coding problem with side information that has received considerable attention over the past decade.
In a general multicast index coding problem, a server/sender wishes to communicate $N^{\prime}$ independent messages to $K^{\prime}$ users through an error-free link. Each client/receiver knows a subset of the $N^{\prime}$ messages and demands a subset of the unknown messages. The server broadcasts packets such that each client can recover the desired messages. The objective is to determine the largest message rate region for a fixed assignment of side information sets. 
If each client demands a single district message, we have a so-called {\it multiple unicast index coding problem.}

For the general index coding problem, an outer bound based on the polymatroidal properties of the entropy function~\cite{networkmatroid} was originally proposed in~\cite[Theorem 3.1]{Lexicographic} and later extended in~\cite[Theorem 1]{onthecapacityindex}. A looser version of~\cite[Theorem 1]{onthecapacityindex} but easier to evaluate for the multiple unicast index coding problem was given in~\cite[Corollary 1]{onthecapacityindex}, which we shall refer to as {\it acyclic outer bound} in the following. 
%
Several inner bounds are known for the index coding problem.
A scheme based on rank minimization of certain matrices was proposed in~\cite{indexcodingwithsi},
and interference alignment based schemes were proposed in~\cite{interferenceali,toplogical}.
%
Since a multiple unicast index coding problem can be represented as a directed graph,
schemes leveraging graph proprieties such as clique-cover, partial clique-cover, local clique cover,  partial local clique covering were proposed in~\cite{chromaticnum,indexcodingwithsi,localgraph,parloccli}, respectively. 
%
Random coding schemes have also been studied. 
The 
schemes proposed in~\cite{indexratedis,indexcorrelated,infortheomulti,roy2016arate} are based on the Heegard-Berger~\cite{heegardberger} idea of source compression with different receiver side information sets. 
By using Slepian-Wolf coding~\cite{slepianwolf}, 
the authors in~\cite{onthecapacityindex} proposed a scheme known as {\it composite coding}, which 
is optimal for the multiple unicast index coding problem with up to five messages.

The index coding problem 
has connection to the coded caching problem as originally formulated by Maddah-Ali and Niensen in~\cite{dvbt2fundamental,decentralizedcoded}, where a server with a library of $N$ files 
is connected via a shared error-free link to $K$ users. 
Each user has a local cache of size $M\leq N$ files to store information.
There are two phases in the caching problem. 
In the placement phase (during network peak-off traffic times) users store parts of the files within their cache without knowledge of later demands. When each user directly copies some bits of the files in his cache, the placement phase is said to be {\it uncoded}; otherwise it is {\it coded}.
If central coordination (among users) during the placement phase is possible, the caching system is said to be {\it centralized}; otherwise it is {\it decentralized}.
In the delivery phase (during network peak traffic times) each user demands a specific file and, based on the users' demands and cache contents, the server broadcasts packets so that each user can recover the demanded file.  
The objective 
is to design a two-phase scheme so that the number of transmitted packets in the delivery phase is minimized for the worst-case demands, referred to as {\it worst-case load}, or just load for simplicity.

The connection between caching and index coding is as follows~\cite{dvbt2fundamental}.
After the users' demands are revealed in a caching scheme with uncoded cache placement, the delivery phase is equivalent to a general index coding problem. Even if the capacity region of the general index coding problem is not known, available inner and outer bounds can be used to bound the worst-case load in the caching problem.
%
To the best of our knowledge, the first outer bound on the worst-case load under the constraint of uncoded cache placement for centralized caching systems was derived in~\cite{ontheoptimality,ourisitinnerbound} and for decentralized caching systems in~\cite{kai-prelimedefense}.
 To this end, we used the acyclic index coding outer bound in~\cite[Corollary 1]{onthecapacityindex} and leveraged the intrinsic symmetries of the caching problem to derive an outer bound that not only outperforms cut-set-based bounds (which are valid for coded cache placement too) but shows the optimality of the Maddah-Ali and Niensen's original 
schemes in~\cite{dvbt2fundamental,decentralizedcoded} for systems with more files than users and under the constraint of uncoded cache placement. 

Our outer bound in~\cite{ontheoptimality,ourisitinnerbound} has been recently shown to be tight for 
caching systems with more users than files as well in~\cite{exactrateuncoded}. The key observation is that certain packets sent in the Maddah-Ali and Niensen's original scheme in~\cite{dvbt2fundamental} are linear combinations of other packets and thus need not be sent. In~\cite{exactrateuncoded}  matching inner and outer bounds for 
systems with uniform demands 
were given.


\paragraph*{Contributions}
This work is motivated by the observation that the coding inner bound is not optimal when applied to the caching problem with uncoded placement. 
We first propose an inner bound for the index coding problem based on 
Han's coding scheme~\cite{hanspaper}, 
Slepian-Wolf coding~\cite{slepianwolf}, and
non-unique decoding~\cite{nonunique}.
This inner bound is proved to strictly improve on composite coding by way of an example.
We then apply the novel inner bound to the caching problem with uncoded cache placement and show that it matches our worst-case load outer bound in~\cite{ontheoptimality,ourisitinnerbound}, thus providing an alternate `source coding with side information' proof to some results in~\cite{exactrateuncoded}. 
{
Compared to the achievable scheme in~\cite{exactrateuncoded}, which is a clever analysis of the linear code originally proposed by Maddah-Ali and Niensen in~\cite{dvbt2fundamental}, our inner bound has the following pleasing features: 
(i) it applies to the general index coding problem,
(ii) it is not restricted to linear codes, and
(iii) it can be easily extended to index coding problems over noisy broadcast channels. 
}


\paragraph*{Paper Outline}
The rest of the paper is organized as follows. 
Section~\ref{sec:model} presents the system models for the index coding and the caching problems, and formally  connects them.
Section~\ref{sec:novel index coding} proves the main result of this paper. 
Section~\ref{sec:conc} concludes the paper.

\paragraph*{Notation}
Calligraphic symbols 
denotes sets.
$|\cdot|$ is 
the cardinality of a set. 
We denote $[1:K]:=\left\{ 1,2,\ldots,K\right\}$ and 
$\mathcal{A\backslash  B}:=\left\{ x\in\Ac|x\notin\mathcal{B}\right\} $.
$\oplus$ represents the bit-wise XOR operation (zeros may need to be appended
to make the vectors have the same length).

\section{System Models and Related Results}
\label{sec:model}

In this section, we start by describing 
the caching problem and
the index coding problem, and 
we finish by discussing their relationship.
By way of an example, we show the need to improve 
the composite coding inner bound 
for the index coding problem 
before inner bounds from index coding can be 
applied to the caching problem.

\subsection{
Caching Problem}
\label{sec:model:cache}

In the coded caching problem 
a central server is equipped with $N$ independent files of $B$ bits.
Files are denoted by $F_{1},\ldots,F_{N}$.
The server is connected to $K$ users through an error-free broadcast link. 

In the placement phase, user $i\in[1:K]$ stores information about the $N$ files in his cache
of size $MB$ bits without knowledge of users' demands. Here $M \in[0,N]$. 
The cache content for user $i\in[1:K]$ is denoted by $Z_i$; we let $\mathbf{Z}:=(Z_{1},\dots,Z_{K})$.
Centralized systems allow for coordination among users in the placement phase, while decentralized systems do not. 
So in decentralized systems the caching functions are   random and independent functions. 

In the delivery phase, each user demands one file and the demand vector $\mathbf{d}:=(d_{1},\dots,d_{K})$ is revealed to everyone, where $F_{d_{i}}, \ d_{i}\in[1:N],$ is the file demanded by user $i\in[1:K]$. 
Given $(\mathbf{Z},\mathbf{d})$, the server broadcasts a message $X_{\mathbf{Z},\mathbf{d}}$ of length $B \, R(\mathbf{d},M)$ bits. It is required that user $i\in[1:K]$ recovers his desired file 
from the broadcast message and his local cache content with arbitrary high probability as $B\to\infty$. 

The objective is to minimize the worst-case load
\begin{align}
R_t^{*}\left(M\right) : =\min \max_{\mathbf{d}} R(\mathbf{d},M),
\label{eq:worst-case load def}
\end{align}
where $t=\textrm{c}$ if the placement phase is centralized, and $t=\textrm{d}$ the placement phase is decentralized.
Note that $R_t^{*}\left(M\right)$ represents the number of transmissions needed to deliver one file to each user.

We briefly revise the details of the scheme originally proposed by Maddah-Ali and Niensen in~\cite{dvbt2fundamental,decentralizedcoded} next.

\subsubsection{Centralized Caching Systems (cMAN)~\cite{dvbt2fundamental}}
Let the cache size be $M=t\frac{N}{K}$, for some positive integer $t\in[0:K]$, and $R[t]$ be the corresponding worst-case load. The worst-case load $R(M)$ for other values of $M$ is obtained as the lower convex envelope of the set of points $\left( t\frac{N}{K}, R[t] \right)$ for  $t\in[0:K]$.

In the placement phase, each file is split into $\binom{K}{t}$ non-overlapping sub-files of equal size. 
The sub-files of $F_{i}$ are denoted by $F_{i,\Wc}$
for $\Wc\subseteq[1:K]$ where $|\Wc|=t$. 
User $k\in[1:K]$ 
fills his cache as
\begin{align}
Z_k 
= \Big( 
F_{i,\Wc}
:  k\in\Wc\subseteq[1:K], |\Wc|=t,  \ i\in[1:N]
\Big).
\label{eq:cMAN cache function}
\end{align}
%
In the delivery phase, the server transmits
\begin{align}
X_{\mathbf{Z},\mathbf{d}} 
= \Big( \oplus_{s\in\Sc}F_{d_{s},\Sc\backslash\{s\}}
: \Sc\subseteq[1:K], |\Sc|=t+1 
\Big),
\label{eq:cMAN tx signal}
\end{align}
which requires broadcasting at a rate
\begin{align}
R_\textrm{cMAN} [t] 
:= \frac{\binom{K}{t+1}}{\binom{K}{t}}.
\end{align}

Let $\mathcal{N}(\mathbf{d})$ be set of distinct demanded files in the demand vector $\mathbf{d}$. 
In~\cite{exactrateuncoded} it was shown that 
among all the $\binom{K}{t+1}$ linear combinations in~\eqref{eq:cMAN tx signal}, $\binom{K-|\mathcal{N}(\mathbf{d})|}{t+1}$ of them can be obtained by linear combinations of the remaining ones and thus need not be transmitted.
Hence, the worst-case load is attained for $|\mathcal{N}(\mathbf{d})|=\min(K,N)$, which requires a broadcast rate of~\cite{exactrateuncoded}
\begin{align}
R_\textrm{c,uncoded placement} [t] 
:= \frac{ \binom{K}{t+1} - \binom{K-\min(K,N)}{t+1} }{\binom{K}{t}}.
\label{eq:centralized optimal}
\end{align}
The worst-case load in~\eqref{eq:centralized optimal} coincides with the outer bound under the constraint of uncoded cache placement in~\cite{ontheoptimality,ourisitinnerbound} and it is thus optimal.

\subsubsection{Decentralized Caching Systems (dMAN)~\cite{decentralizedcoded}} 
In decentralized systems, user coordination during the placement phase is not possible,
so each user stores a subset of $\frac{M}{N}B$ bits of each file, chosen uniformly and independently at random. 
Given the cache content of all the users, the bits of the files can be grouped into sub-files $F_{i,\Wc}$, where $F_{i,\Wc}$ is the set of bits of file $i\in[1:N]$ that are only known by the users in $\Wc\subseteq[1:K]$. 
By the Law of Large Numbers, the size of the sub-files converges in probability to
\begin{align}
\frac{|F_{i,\Wc}|}{B} \stackrel{p.}{\to}  \left(\frac{M}{N}\right)^{|\Wc|}   \left(1-\frac{M}{N}\right)^{K-|\Wc|}
          \textrm{when $B\to\infty$}. 
\label{eq:law of large number}
\end{align}
In the delivery phase, for each $t\in [0:K-1]$, 
all the $\binom{K}{t+1}$ sub-files $F_{i,\Wc}$ with $|\Wc|=t$ and $i\in[1:N]$ are gathered together; since they all 
have approximately the same length that only depends on how many users 
have stored them in their cache
(given by~\eqref{eq:law of large number}), the 
server uses the {cMAN} scheme 
for $M=t\frac{N}{K}$ to deliver them. Thus, the worst-case load of the {dMAN} scheme is 
\begin{align}
R_\textrm{dMAN} \left(M\right) &:=\sum_{t\in [0:K-1]} \binom{K}{t+1} \left(\frac{M}{N}\right)^{t}\left(1-\frac{M}{N}\right)^{K-t}\notag\\
&=\frac{1-\frac{M}{N}}{\frac{M}{N}}\left[1-\left(1-\frac{M}{N}\right)^{K}\right].
\end{align}
The optimal load for decentralized caching systems with uncoded cache placement can be achieved following the {dMAN} original idea without the redundant transmissions in the underlying {cMAN} scheme, which leads to~\cite{exactrateuncoded}
\begin{align}
R_\textrm{d,uncoded placement} \left(M\right) := \frac{1-\frac{M}{N}}{\frac{M}{N}} \left[1-\left(1-\frac{M}{N}\right)^{\min(K,N)}\right].
\label{eq:decentralized optimal}
\end{align}
The worst-case load in~\eqref{eq:decentralized optimal} coincides with the outer bound under the constraint of uncoded cache placement in~\cite{kai-prelimedefense} and it is thus optimal.

Before we connect the caching problem with uncoded cache placement to the index coding problem, we need to introduce the index coding problem formally.


\subsection{Index Coding Problem}
\label{sec:model:IC}

In the index coding problem 
a central server with $N^{\prime}$ independent messages is connected to $K^{\prime}$ users.
Each user $j\in [1:K^{\prime}]$ demands a set of messages indexed by $\Dc_{j} \subseteq [1:N^{\prime}]$ and knows a set of messages indexed by $\Ac_{j} \subseteq [1:N^{\prime}]$. 
In order to avoid trivial problems, it is assumed that $\Dc_{j} \not= \emptyset$, $\Ac_{j} \not= [1:N^{\prime}]$, and $\Dc_{j}\cap\Ac_{j}=\emptyset$.  
The server is connected to the users through a noiseless channel with alphabet $\mathcal{X}$. Without loss of generality we can take $\mathcal{X}$ to be GF($2$)~\cite{onthecapacityindex}.
%
A $(2^{nR_{1}},\ldots,2^{nR_{N^{\prime}}},n,\epsilon_n)$-code 
for this index coding problem  is defined as follows.

Each message $M_{i}$, for $i\in [1:N^{\prime}],$ is uniformly distributed in 
$[1:2^{nR_{i}}]$ 
where $n$ is the block-length, $R_{i} \geq 0$ is the transmission rate in bits per channel use.
In order to satisfy users' demands, the server broadcasts $X^{n}=\mathsf{enc}(M_{1},\ldots,M_{N^{\prime}})\in \mathcal{X}^{n}$ where $\mathsf{enc}$ is the encoding function.
Each user $j\in[1:K^{\prime}]$ estimates the messages indexed by $\Dc_{j}$ by the decoding function $\mathsf{dec}_j\big( X^{n}, (M_{i}:i\in\Ac_{j}) \big)$. The probability of error is
\begin{align*}
\epsilon_n := \max_{j\in[1:K^{\prime}]}\Pr\left[\mathsf{dec}_j\big( X^{n}, (M_{i}:i\in\Ac_{j}) \big) \not= (M_{i}:i\in\Dc_{j}) \right]. 
\end{align*}
A rate vector $(R_{1},\ldots,R_{N^{\prime}})$ is said to be achievable if there exists a family of 
$(2^{nR_{1}},\ldots,2^{nR_{N^{\prime}}},n,\epsilon_n)$-codes 
with $\lim_{n\to\infty}\epsilon_n=0$. 
%

For later use, we close this subsection with a description of the composite coding inner bound, which was proposed for the multiple unicast index coding problem in~\cite{onthecapacityindex}. We trivially extended it here to the general index coding problem.

\subsubsection*{Composite Coding Inner Bound}
Composite coding is a two-stage scheme  based on binning and non-unique decoding.
In the first encoding stage, for each $\Jc\subseteq[1:N^{\prime}]$, the messages $(M_{i} : i\in\Jc)$ are encoded into the `composite index' $W_{\Jc}\in[1:2^{nS_{\Jc}}]$ 
based on random binning at some rate $S_{\Jc} \geq 0$.  
By convention $S_{\emptyset}=0$. 
In the second encoding stage, 
the collection of all composite indices $(W_{\Jc} : \Jc\subseteq[1:N^{\prime}])$ is mapped into a length-$n$ 
sequence $X^{n}\in \mathcal{X}^{n}$. 
In the first decoding stage, every user 
recovers all composite indices by making use of the available side information.
In the second decoding stage, user $j\in[1:K^{\prime}]$ chooses a set $\Kc_{j}$ such that $\Dc_{j} \subseteq  \Kc_{j} \subseteq [1:N^{\prime}]\backslash\Ac_{j}$ 
and simultaneously decodes all messages $(M_{i} : i\in\Kc_{j})$, 
based on the recovered $(W_{\Jc} : \Jc\subseteq \Kc_{j}\cup\Ac_{j})$.
The achievable rate region with composite coding is as follows.
\begin{thm}[Composite Coding Inner Bound, generalization of~\cite{onthecapacityindex} to allow for multicast messages]
\label{thm2 composite coding}
\begin{subequations} 
A non-negative rate tuple $\mathbf{R} :=(R_{1},\ldots,R_{N^{\prime}})$ is achievable
for the index coding problem $\Big( (\Ac_{j},\Dc_{j}) : j\in[1:K^{\prime}] \Big)$
with $N^{\prime} = \left| \cup_{j\in[1:K^{\prime}]} \Ac_{j}\cup\Dc_{j} \right|$
if
\begin{align}
&\mathbf{R}\in
\bigcap_{j\in [1:K^{\prime}]} \quad
\bigcup_{\small
\Kc_{j}:
\Dc_{j}\subseteq \Kc_{j}\subseteq [1:N^{\prime}]\backslash \Ac_{j}
}
\mathscr{R}_\text{cc}(\Kc_{j}|\Ac_{j},  \Dc_{j}),
\label{eq:composite 1}
\\
&\mathscr{R}_\text{cc}(\Kc|\Ac,\Dc)
:=  \bigcap_{\Jc : \Jc\subseteq\Kc} 
\left\{ \sum_{i\in\Jc}R_{i} < v_{\Jc} \right\},
\label{eq:composite 2}
\end{align}
where in~\eqref{eq:composite 2} $v_{\Jc}$ is defined as
\begin{align}
v_{\Jc}:=\sum_{\Pc: \Pc\subseteq\Ac\cup\Kc , \Pc\cap\Jc\neq\emptyset} 
S_{\Pc}, %
\label{eq:composite 3 vJ}
\end{align} 
and where in~\eqref{eq:composite 3 vJ} the non-negative quantities $(S_{\Jc}:\Jc\subseteq[1:N])$ must satisfy 
\begin{align}
\sum_{\Jc:\Jc\in[1:N^{\prime}], \Jc\nsubseteq\Ac_{j}}S_{\Jc}\leq  \log_2(|\mathcal{X}|), 
\quad \forall j\in [1:K^{\prime}].
\label{eq:composite 4 decompression}
\end{align}
\end{subequations} 
\end{thm}
Note that the constrain in~\eqref{eq:composite 4 decompression} is from the first decoding stage
and the region $\mathscr{R}_\text{cc}(\Kc_{j}|\Ac_{j},  \Dc_{j})$ in~\eqref{eq:composite 1} is from the second decoding stage at receiver $j\in[1:K^{\prime}]$.

\subsection{Connecting Caching to Index Coding}
Under the constraint of uncoded cache placement, 
when the cache contents and the demands are fixed, 
the delivery phase of the caching problem is equivalent to the following index coding problem. 
For each $i\in\mathcal{N}(\mathbf{d})$ and for each $\Wc\subseteq[1:K]$, 
the sub-file $F_{i,\Wc}$ (containing the bits of file $F_{i}$ within the cache of the users indexed by $\Wc$) is an independent message in the index coding problem with user set $[1:K]$
Hence, by using the notation introduced in Sections~\ref{sec:model:IC} and~\ref{sec:model:cache},
$K^{\prime} = K$ and 
$N^{\prime} = |\mathcal{N}(\mathbf{d})| (2^K-1)$. 
For each user $k\in [1:K]$ in this general index coding problem, 
the desired message and side information sets are 
\begin{align}
\Dc_{k}&= \big(F_{d_{k},\Wc} : \Wc\subseteq[1:K], k\notin\Wc \big), 
\label{eq:dk def for caching}
\\
\Ac_{k}&= \big(F_{i,\Wc} : \Wc\subseteq[1:K], i\in\mathcal{N}(\mathbf{d}), k\in \Wc \big).
\label{eq:ak def for caching}
\end{align}

In~\cite{ontheoptimality,ourisitinnerbound}, we proposed an outer bound on the worst-case load in centralized caching systems under the constraint of uncoded cache placement by exploiting the acyclic index coding outer bound in~\cite[Corollary 1]{onthecapacityindex}. For a demand vector $\mathbf{d}$, we considered all possible multiple unicast index coding problems with $|\mathcal{N}(\mathbf{d})|$ users. By summing together the resulting bounds and by taking the worst-case demand vector $\mathbf{d}$, we showed that~\eqref{eq:centralized optimal} is a lower bound to the worst-case load under uncoded cache placement for centralized systems~\cite{ontheoptimality,ourisitinnerbound}.
We followed a similar approach for decentralized caching systems in~\cite{kai-prelimedefense}.

When we attempted to match the worst-case load lower bounds in~\eqref{eq:centralized optimal} and~\eqref{eq:decentralized optimal}
with an achievable load from the composite coding inner bound for index coding  
in Theorem~\ref{thm2 composite coding} we failed\footnote{
The reason why we 
do not consider the other index coding achievable schemes we mentioned in the Introduction is because they do not provide easily computable rate expressions for the general index coding problem, 
or because they were designed for the case of two messages only. 
}. 
The following example shows that composite coding is insufficient for the index coding problem.
This was already pointed out in~\cite{onthecapacityindex}. The example we give next will be used later on to show that our proposed index coding inner bound is strictly better than composite coding.
\begin{example}
\rm
\label{ex:Insufficiency on Composite Coding}
Consider a multiple unicast index coding problem with $K=6$ equal rate messages and with
\begin{align*}
& \Dc_{1}=\{1\}, \quad \Ac_{1}=\{3,4\},\\
& \Dc_{2}=\{2\}, \quad \Ac_{2}=\{4,5\},\\
& \Dc_{3}=\{3\}, \quad \Ac_{3}=\{5,6\},\\
& \Dc_{4}=\{4\}, \quad \Ac_{4}=\{2,3,6\},\\
& \Dc_{5}=\{5\}, \quad \Ac_{5}=\{1,4,6\},\\
& \Dc_{6}=\{6\}, \quad \Ac_{6}=\{1,2\}.
\end{align*}

Composite Coding Inner Bound.
In~\cite[Example~1]{liu2017ondistributed} the authors showed that the largest symmetric rate with the composite coding inner bound in Theorem~\ref{thm2 composite coding} for this problem is $R_\text{sym,cc}=0.2963 \cdot \log_2(|\mathcal{X}|)$. It the same paper, the authors proposed an extension of the composite coding idea (see~\cite[Section III.B]{liu2017ondistributed}) and showed that this extended scheme for this problem gives $R_\text{sym,enhanced cc}=0.2987\cdot \log_2(|\mathcal{X}|)$.

Converse.
Give message $F_5$ as additional side information to receiver~1 so that the new side information set satisfied $\{3,4,5\} \subset \Ac_{2}$.
With this receiver~1, in addition to message~1, can decode message~2  and then message~6.
Thus 
\begin{align}
3R_\text{sym} \leq \lim_{n\to\infty}\frac{1}{n}H(X^n) \leq  \log_2(|\mathcal{X}|).
\label{eq:outer example} 
\end{align}
Next we show that $R_\text{sym}=1/3\cdot \log_2(|\mathcal{X}|)$ is tight. 
This shows the strict sub-optimality of composite coding and its extension.

Achievability.
Take the messages to be binary digits.
All users can be satisfied by the transmission of the three coded bits
$X=(F_1\oplus F_3\oplus F_4, \ F_2\oplus F_4\oplus F_5, \  F_1\oplus F_2\oplus F_6)$.
Receivers~1,~2 and~6 can `read off' the desired message bit from one of the transmitted bits after subtracting the known bits.
Receiver~3 first sums the three transmitted bits and then recovers $F_3$ thanks to its side information; receivers~4 and~5 proceed similarly. This shows that one bit per user can be delivered in one channel use, where one channel use corresponds to three bits. Therefore, $R_\text{sym}=1/3\cdot \log_2(|\mathcal{X}|)$ is achievable and is optimal.
\hfill$\square$
\end{example}

Given that composite coding is insufficient, in the rest of the paper we derive a novel index coding achievable scheme, which we shall prove to strictly improve on composite coding and to be sufficient for caching.

\section{Novel Index Coding Scheme and its Application to the Caching Problem}
\label{sec:novel index coding}
\subsection{Novel Index Coding Scheme}
\label{sec:novel IC scheme}
In this section, we first introduce a novel  achievable scheme for index coding
and then prove that 
it strictly outperforms composite coding by continuing Example~\ref{ex:Insufficiency on Composite Coding}.
%
Intuitively, the improvements in our scheme come from:
\begin{itemize}
\item
For each subset $\Jc\subseteq [1:K^{\prime}]$ in the composite coding scheme, the composite index $W_{\Jc}$ is determined by the messages indexed by $\Jc$. Thus, composite indices are correlated among themselves. We leverage this correlation to lower the required rate in the first decoding stage.
\item 
In the composite coding scheme, 
decoder $j \in[1:K^{\prime}]$ wants to recover uniquely the messages in $\Kc_{j}$, and for that he only uses the composite indices $(W_{\Jc} : \Jc \subseteq \Kc_{j}\cup \Ac{j})$. In our proposed scheme, every  user  uses all the composite messages $(X_{\Jc} : \Jc \subseteq [1:N'])$ to uniquely recover the desired messages in $\Dc_{j}$ and non-uniquely those in $\Kc_{j}\backslash \Dc_{j}$, while the remaining messages are treated as noise.
\end{itemize}


\begin{thm}[Novel Achievable Scheme for Index Coding]
\label{thm3 novel index coding} 
\begin{subequations}
A non-negative rate tuple $\mathbf{R} :=(R_{1},\ldots,R_{N^{\prime}})$ is achievable
for the index coding problem $\Big( (\Ac_{j},\Dc_{j}) : j\in[1:K^{\prime}] \Big)$
with $N^{\prime} = \left| \cup_{j\in[1:K^{\prime}]} \Ac_{j}\cup\Dc_{j} \right|$ if 
\begin{align}
&\mathbf{R}\in
\bigcap_{j\in [1:K^{\prime}]} \quad
\bigcup_{\small
\Kc_{j}:
\Dc_{j}\subseteq \Kc_{j}\subseteq [1:N^{\prime}]\backslash \Ac_{j}
}
\mathscr{R}(\Kc_{j}|\Ac_{j},  \Dc_{j}),
\label{eq:novel 1}
\\
&\mathscr{R}(\Kc|\Ac,\Dc)
:= \bigcap_{\Jc : \Jc\subseteq\Kc,  \Dc\cap\Jc\not=\emptyset}
\left\{ \sum_{i\in\Jc}R_{i} < \kappa_{\Jc} \right\},
\label{eq:novel 2}
\end{align}
where in~\eqref{eq:novel 2} $\kappa_{\Jc}$ is defined as
\begin{align}
\kappa_{\Jc}:=
&I\Big(
  \big(U_{i} : i\in\Jc\big) \ ; \ \big(X_{\Pc}: \Pc\subseteq[1:N^{\prime}] \big)
\notag\\&\qquad
\Big|
\big(U_{i} : i\in\Ac_{j} \cup \Kc_{j}\setminus\Jc \big)
\Big),
\label{eq:wJ}
\end{align}
for 
some independent auxiliary random variables $(U_{i} : i\in[1:N^{\prime}])$ and
some functions $\Big(f_{\Pc} : \Pc\subseteq[1:N^{\prime}] \Big)$, 
such that $X_{\Pc}=f_{\Pc}\Big(\big(U_{i}:i\in\Pc\big)\Big)$ and satisfying for all $j\in [1:K^{\prime}]$
\begin{align}
H\Big(
\big(X_{\Pc}:  \Pc\subseteq[1:N^{\prime}]\big) 
\big|
\big(U_{i}:i\in\Ac_{j}\big)
\Big) \leq  \log_2(|\mathcal{X}|). 
\label{eq:H(X)<c}
\end{align} 
\end{subequations}
\end{thm}
\begin{IEEEproof}  
Intuitively, the proof is as follows.

Encoding.
Each message $M_i, i\in[1:N^{\prime}],$ is encoded into a codeword $U_{i}^{n}$ 
generated in an i.i.d. fashion according to some distribution $p_{U_{i}}$.
Then the collection $(U_{i}^{n} : i\in\Pc)$ is mapped into a `composite index' $X_{\Pc}^{n}\in[1:2^{n S_{\Pc}}]$, for all $\Pc\subseteq[1:N^{\prime}]$, by using the function $f_{\Pc}$ component-wise. 
Each receiver observes the `channel input' $X^{n}:={\sf bin}\big(X_{\Pc}^{n} : \Pc\subseteq[1:N^{\prime}] \big) \in\mathcal{X}^n$, where ${\sf bin}$ is the bin index of the collection 
$\big(X_{\Pc}^{n} : \Pc\subseteq[1:N^{\prime}] \big)$. 
Binning is done uniformly and independently. 

Decoding.
Receiver $j\in [1:K^{\prime}]$, given the side information $\Ac_{j}$, can recover the `channel input' $X^{n}$ if the condition in~\eqref{eq:H(X)<c} is satisfied, i.e., only the `composite indices' that are not fully determined by the side information must be recovered. 
Finally, receiver $j\in [1:K^{\prime}]$ chooses a set $\Kc_{j}\in [1:N^{\prime}]$ such that
it includes all the desired messages but none of the side information messages (that is, $\Dc_{j} \subseteq  \Kc_{j} \subseteq [1:N^{\prime}]\backslash\Ac_{j}$);
he then simultaneously decodes all messages $(M_{i} : i\in\Kc_{j})$, but uniquely only the messages in $\Dc_{j}$.
For this equivalent multiple access channel with user set $\Kc_{j}$, the achievable region is in the form of~\eqref{eq:novel 1} where the messages indexed by $\Jc$  can be reliably decoded, given that those in the side information or already decoded are known (that is, given the messages indexed by $\Ac_{j} \cup \Kc_{j}\setminus\Jc$), if the condition in~\eqref{eq:wJ} is satisfied.

This concludes the proof.
\end{IEEEproof}

\begin{cor}
\label{cor outperform composite coding}
The composite coding region in Theorem~\ref{thm2 composite coding}
is a special case of our Theorem~\ref{thm3 novel index coding}.
\end{cor}
\begin{IEEEproof}
In general, for a set $\Bc\subseteq[1:N^{\prime}]$ and for the auxiliary random variables as defined in Theorem~\ref{thm3 novel index coding}, we have
\begin{align}
&
H\Big( \big(X_{\Pc}:\Pc\subseteq[1:N^{\prime}]\big) 
\Big|
\big(U_{i} : i\in \Bc\big) 
\Big)
\notag\\&
\leq H\Big( \big(X_{\Pc}:\Pc\subseteq[1:N^{\prime}], \Pc\not\subseteq \Bc \big) 
\Big)
\notag\\&\leq 
\sum_{\Pc:\Pc\subseteq[1:N^{\prime}] , \Pc\not\subseteq \Bc}
H\big( X_{\Pc} \big)
\notag\\&\leq 
\sum_{\Pc:\Pc\subseteq[1:N^{\prime}] , \Pc\not\subseteq \Bc}
S_{\Pc}, \ \text{ where $\log_{2}(|\mathcal{X}_{\Pc}|)=S_{\Pc}$}. 
\label{eq:composite ineqs}
\end{align}

In the following, we choose $(U_{i}: i\in [1:N^{\prime}])$ and $(X_{\Pc}: \Pc\subseteq[1:N^{\prime}])$ such that all the inequality leading to~\eqref{eq:composite ineqs} holds with equality for any $\Bc\subseteq[1:N^{\prime}]$, that is, we construct random variables $(X_{\Pc}: \Pc\subseteq[1:N^{\prime}])$ that are independent and uniformly distributed, where the alphabet of $X_{\Pc}$ has support of size $|\mathcal{X}_{\Pc}|=2^{S_{\Pc}}$.
With this choice of auxiliary random variables we show that Theorem~\ref{thm3 novel index coding} reduces to Theorem~\ref{thm2 composite coding}.

Assume that $S_{\Pc} \log_2(|\mathcal{X}|)$ is an integer for all $\Pc\subseteq[1:N^{\prime}]$.
Let $U_{i}$, for $i\in [1:N^{\prime}]$, be an equally likely binary vector of length $L_{i}$. 
Let $X_{\Pc}$ be a binary vector of length $S_{\Pc} \log_2(|\mathcal{X}|)$ obtained as a linear code for the collection of bits in $(U_{i}, i\in\Pc)$.
If $L_{i}\geq\sum_{\Pc\subseteq[1:N^{\prime}]:i\in\Pc}S_{\Pc}\log_{2}(|\mathcal{X}|)$ for all $i\in [1:N^{\prime}]$, then all the linear combinations that determine $X_{\Pc}$ can be chosen to be independent and therefore all the inequalities leading to~\eqref{eq:composite ineqs} holds with such choice of auxiliary random variables.
As a result, 
we have that
the bound in~\eqref{eq:H(X)<c} reduces to the one in~\eqref{eq:composite 4 decompression} 
by using~\eqref{eq:composite ineqs} with $\Bc=\Ac_{j}$, and that
the bound in~\eqref{eq:wJ} reduces to the one in~\eqref{eq:composite 3 vJ}
by using~\eqref{eq:composite ineqs} twice, once with $\Bc=\Ac \cup \Kc\setminus\Jc$ and once with $\Bc=\Ac \cup \Kc$, which is so because
\begin{align*}
\kappa_{\Jc}
&=\sum_{\Pc:\Pc\subseteq[1:N^{\prime}] : \Pc\not\subseteq (\Ac \cup \Kc\setminus\Jc)} \!\!\!\!S_{\Pc}
 -\sum_{\Pc:\Pc\subseteq[1:N^{\prime}] : \Pc\not\subseteq (\Ac \cup \Kc)} \!\!\!\!S_{\Pc}
\notag
\\
&=\sum_{\Pc:\Pc\subseteq \Ac \cup \Kc : \Pc\cap\Jc\not=\emptyset} \!\!\!\!S_{\Pc}.
\end{align*}

This concludes the proof.
\end{IEEEproof}

\begin{example}
\rm
\label{ex:Insufficiency on Composite Coding cont}
We continue Example~\ref{ex:Insufficiency on Composite Coding}.
Let each file be an independent bit, 
$\Kc_{j}=\Dc_{j}$ for $j\in [1:6]$, and 
\begin{align*}
  &U_{1} = F_1, \ U_{2} = F_2, \cdots, U_{6} = F_6,
\\&\text{for all $\Pc\subseteq[1:6]$ set $X_{\Pc}=0$ except}
\\&X_{\{1,3,4\}}=U_{1}\oplus U_{3}\oplus U_{4},
\\&X_{\{2,4,5\}}=U_{2}\oplus U_{4}\oplus U_{5},
\\&X_{\{1,2,6\}}=U_{1}\oplus U_{2}\oplus U_{6},
\\&X=(X_{\{1,3,4\}},X_{\{2,4,5\}},X_{\{1,2,6\}}).
\end{align*}
Here $\mathcal{X}=\text{GF($2^3$)}$ so one channel use corresponds to three bits.
From~\eqref{eq:wJ}, we have that for example the rate of user~5 is bounded by
\begin{align*}
R_\text{sym} 
&\leq 
I(
U_{5}
;
U_{1}\oplus U_{3}\oplus U_{4},
U_{2}\oplus U_{4}\oplus U_{5},
U_{1}\oplus U_{2}\oplus U_{6}
|
U_{1}, U_{4}, U_{6}
)
\\
&= 
I(
U_{5}
;
U_{3},
U_{2}\oplus U_{5},
U_{2}
)
= 
I(
U_{5}
;
U_{2},
U_{3},
U_{5}
)
\\
&= 
I(
U_{5}
;
U_{5}
)
=
H (U_{5})=1/3\cdot \log_2(|\mathcal{X}|), 
\end{align*}
and similarly for all the other users.
As a result, $R_\text{sym}= 1/3\cdot \log_2(|\mathcal{X}|)$ is achievable by the proposed scheme and coincides with the outer bound.
\hfill$\square$
\end{example}

\subsection{Application to the Caching Problem}
\label{sec:extension to caching}
We are now ready to show that Theorem~\ref{thm3 novel index coding} can be used to determine the optimal load in caching problems under the constraint of uncoded cache placement.

\begin{thm}
\label{thm:match centralized} 
For a caching system under the constraint of uncoded cache placement,  Theorem~\ref{thm3 novel index coding} achieves the worst-case loads in~\eqref{eq:centralized optimal} and~\eqref{eq:decentralized optimal} for centralized and decentralized caching systems, respectively.
\end{thm}
\begin{IEEEproof}
We only do the proof for centralized caching systems under the constraint of uncoded cache placement as the one for  decentralized systems follows similarly.

We use the same placement phase as {cMAN} for $M=t\frac{N}{K}$, for $t\in[0:K]$, so that the delivery phase is equivalent to an index coding problem with $K$ users in which each sub-file $F_{i,\Wc}$, for $i\in \mathcal{N}(\mathbf{d})$, $\Wc\subseteq[1:K]$ and $|\Wc|=t$, is an independent message, 
and where the desired message and side information sets are given by~\eqref{eq:dk def for caching} and~\eqref{eq:ak def for caching}, respectively.
Note that the message rates in this equivalent index coding problem are identical by construction and the number of messages for the worst case-load is $N^{\prime} = \min(N,K) \binom{K}{t}$.
%

In Theorem~\ref{thm3 novel index coding}, following in Example~\ref{ex:Insufficiency on Composite Coding cont},
we let $\Kc_{j}=\Dc_{j}$ for $j\in [1:K]$, 
we represent $F_{i,\Wc}$ as a binary vector for length $k$ and we let the corresponding random variable $U$ to be equal to the message. 
We also let $X_{\Pc}$ to be non zero only for the linear combinations of messages sent by the scheme in~\cite{exactrateuncoded}. With this we have 
$R_\text{sym} = H(U) = k$ and
$\log_2(|\mathcal{X}|) = H(X) = k\left(\binom{K}{t+1}-\binom{K-\min(N,K)}{t+1}\right)$,
so the symmetric rate is
\begin{align*}
R_\text{sym} =  \frac{1}{\binom{K}{t+1}-\binom{K-|\mathcal{N}(\mathbf{d})|}{t+1}}\log_2(|\mathcal{X}|).
\end{align*}

Each receiver in the original caching problem is interested in recovering $\binom{K}{t}$ messages, 
or one file of $k\binom{K}{t}$ bits,
thus the `sum-rate rate' delivered to each user is
\begin{align*}
R_\text{sum-rate} = \frac{\binom{K}{t}}{\binom{K}{t+1}-\binom{K-|\mathcal{N}(\mathbf{d})|}{t+1}}\log_2(|\mathcal{X}|) \ 
\left[\frac{\text{bits}}{\text{ch.use}}\right].
\end{align*}
The load in the caching problem is the number of transmissions (channel uses) needed to deliver one file to each user, thus the inverse of $R_\text{sum-rate}$ for $|\mathcal{X}|=2$ indeed corresponds to the load in~\eqref{eq:centralized optimal}.

\end{IEEEproof}

\section{Conclusion}
\label{sec:conc}
In this paper, we investigated the index coding problem and its application to the caching problem with uncoded placement. We proposed a novel index coding inner bound based on distributed source coding that provably strictly improves on composite coding. The novel index coding scheme was then shown to be sufficient to match a known outer bound on the optimal worst-case load in caching systems under the constraint of uncoded cache placement. 
%

\section*{Acknowledgment}
The work of K. Wan and D. Tuninetti is supported by Labex DigiCosme and in part by NSF 1527059, respectively

\newpage
\bibliographystyle{IEEEtran}
\bibliography{IEEEabrv,IEEEexample}
\end{document}